\documentclass[runningheads]{llncs}
\usepackage[T1]{fontenc}
\usepackage{graphicx,verbatim}
\usepackage{amsfonts}
\usepackage{bm}
\usepackage{algorithm, algpseudocode}
\usepackage{booktabs}
\usepackage{colortbl}%
  \newcommand{\myrowcolor}{\rowcolor[gray]{0.925}}
\usepackage{amsmath}
\usepackage{subcaption}
\usepackage{hyperref}

\begin{document}
\title{Tackling Hallucination from Conditional Models for Medical Image Reconstruction with DynamicDPS}
%

\author{Seunghoi Kim$^*$\inst{1,2}, Henry F. J. Tregidgo$^*$\inst{1,2}, Matteo Figini\inst{1,3}, Chen Jin\inst{5}, Sarang Joshi\inst{4}, Daniel C. Alexander\inst{1,3} }  
\authorrunning{Kim et al.}
\institute{%
Hawkes Institute, University College London, UK\and
Department of Medical Physics and Biomedical Engineering, University College London, UK \and
Department of Computer Science, University College London, UK\\ \and
The University of Utah, USA, 
\and
Centre for AI, DS\&AI, AstraZeneca, UK
}

\maketitle              

\setcounter{footnote}{0}
\def\thefootnote{*}\footnotetext{These authors contributed equally to this work \\
Corresponding author E-mail: seunghoi.kim.17@ucl.ac.uk}\def\thefootnote{\arabic{footnote}}

\begin{abstract}
Hallucinations are spurious structures not present in the ground truth, posing a critical challenge in medical image reconstruction, especially for data-driven conditional models. 
We hypothesize that combining an unconditional diffusion model with data consistency, trained on a diverse dataset, can reduce these hallucinations.
Based on this, we propose DynamicDPS, a diffusion-based framework that integrates conditional and unconditional diffusion models to enhance low-quality medical images while systematically reducing hallucinations.
Our approach first generates an initial reconstruction using a conditional model, then refines it with an adaptive diffusion-based inverse problem solver. DynamicDPS skips early stage in the reverse process by selecting an optimal starting time point per sample and applies Wolfe’s line search for adaptive step sizes, improving both efficiency and image fidelity. 
Using diffusion priors and data consistency, our method effectively reduces hallucinations from any conditional model output.
We validate its effectiveness in Image Quality Transfer for low-field MRI enhancement. 
Extensive evaluations on synthetic and real MR scans, including a downstream task for tissue volume estimation, show that DynamicDPS reduces hallucinations, improving relative volume estimation by over 15\% for critical tissues while using only 5\% of the sampling steps required by baseline diffusion models. 
As a model-agnostic and fine-tuning-free approach, DynamicDPS offers a robust solution for hallucination reduction in medical imaging. Code is available at \url{https://github.com/edshkim98/DynamicDPS}.



\end{abstract}

\section{Introduction}
A key impediment to the adoption of generative models for Magnetic Resonance Imaging (MRI) reconstruction and enhancement is the possibility of \emph{hallucinations}, spurious yet realistic-looking structures not present in the imaging target. 
Such errors can distort downstream analyses, limiting clinical applicability in critical settings. 
Hence, we present a novel framework for addressing hallucinations from conditional models while minimizing additional training and inference costs. 
While broadly applicable to inverse problems, we focus on low-field MRI enhancement via Image Quality Transfer (IQT), using the term “reconstruction” in the broader sense throughout the paper.


\begin{figure}[!tbp]
\centering
\includegraphics[width=0.9\textwidth]{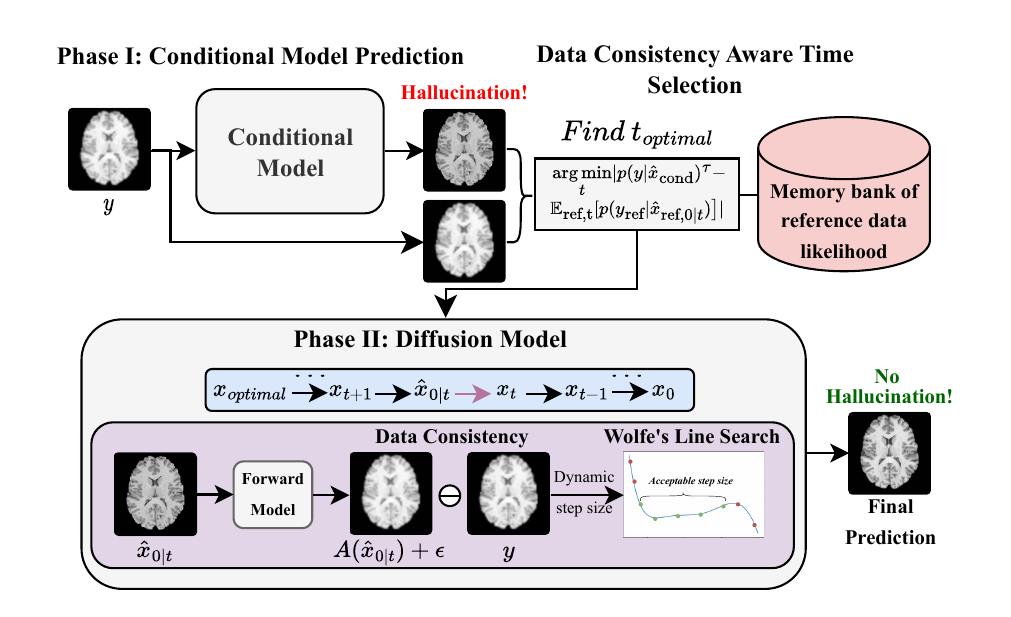}
\caption{The overview of our proposed framework. The framework first predicts a high-quality image using a conditional model. Then, it calculates optimal starting time point in the reverse process via Data Consistency Aware Time Selection (DCATS). At last, the prediction is refined using DynamicDPS that adaptively adjusts step size in each time step using Wolfe's line search.}
\label{fig:method_overview}
\end{figure}

IQT \cite{iqt_pio} is a machine learning framework combining aspects of super-resolution, de-noising and contrast enhancement, enhancing visible tissue properties in low-quality images to the equivalent high-quality counterpart. 
IQT has been applied using various approaches, ranging from classical regression and random forests to deep learning architectures \cite{iqt_stochastic,iqt_uncertainty}, contrast-agnostic models \cite{synth_survey,synthsr}, attention-based CNNs \cite{uliqt}, and 3D conditional diffusion-based methods \cite{diffusioniqt} that work robustly under heavily under-sampled~MRI. This contrasts to super-resolution work~\cite{srcnn,sr_mrigan,sr_soupgan,segsrgan,sr_transformer,dip} which focus more directly on the resolution problem only.

Recent work \cite{localdiff,patchood,modeinterpolation,distributionshift} has shown that hallucinations typically  arise in image regions that are \emph{out-of-distribution (OOD)} for the conditional model being applied.
To date relatively little work has appeared on quantifying hallucinations in medical imaging~\cite{hallucination_tomo,hallucination_metric}, and efforts to minimize hallucinations from conditional generative models have been limited to handling  OOD regions separately to \emph{in distribution (IND)} regions~\cite{localdiff}. 
However, we hypothesize that unconditional diffusion models for inverse problems (or score-matching networks) \cite{chung2023diffusion,improving_inverse,song_score,zeroddrm,diffpir,blinddps} offer a framework for hallucination reduction by combining rich domain-specific priors, learned from high-quality medical datasets without being conditioned on specific measurements with explicit data consistency optimization. The rationale behind this hypothesis is further detailed in Sec~\ref{sec:hypothesis}.

One example, Diffusion Posterior Sampling (DPS)~\cite{chung2023diffusion}, adopts a Bayesian framework, using the score-matching network as strong domain prior and likelihood estimation for data consistency in the reverse process to provide feasible solutions for given measurements. 
However, this approach typically requires a large number of sampling steps at inference time (e.g. T=1000).
Similarly, Lin et al.~\cite{zeroshotmri} extended DPS to low-field MRI enhancement, but incorporated test-time model-parameter optimization, increasing inference cost.

In this work, we propose \emph{DynamicDPS}, a novel framework for MRI reconstruction that extends DPS~\cite{chung2023diffusion} to mitigate hallucinations produced by conditional models. 
DPS offers a principled framework for hallucination reduction, but its high inference cost and fixed step sizes limit practicality and can lead to sub-optimal convergence. Our approach addresses these by improving efficiency and robustness, enabling hallucination mitigation at lower computational cost.

Figure~\ref{fig:method_overview} provides an overview of our method. A conditional model first generates a high-field-like prediction as a prior. DCATS selects an optimal intermediate starting point to reduce diffusion steps, after which the prediction is refined via score-matching and data consistency optimization. We dynamically adjust the data-consistency step size to suppress hallucinations and accelerate sampling by skipping unnecessary steps. Our model-agnostic approach integrates seamlessly with existing conditional models. Applied to low-field MRI enhancement via IQT, DynamicDPS reduces hallucinations and achieves over 80\% faster inference than prior diffusion-based methods. To the best of our knowledge, this is the first work to explicitly address hallucinations in MRI reconstruction.

\section{Method}
We propose DynamicDPS, a novel framework to mitigate hallucinations. 
This section provides a detailed overview of our approach and how it addresses the limitations.
We first formulate the problem of IQT and hallucinations, followed by our hypothesis and theoretical justification for tackling these challenges.
Finally, we present a detailed breakdown of each component within our framework.

\subsection{Problem Formulation}
\label{sec:problem_formulation}

\emph{Image Quality Transfer (IQT)}~\cite{iqt_pio} aims to learn a mapping function between low-field (LF) images \(\mathbf{y} \in \mathbb{R}^M\) and their high-field (HF) counterparts \(\mathbf{x} \in \mathbb{R}^N\). 
However, due to the need of paired data training, most networks~\cite{iqt_pio,iqt_stochastic,diffusioniqt} simulate LF images using a degradation model and train in a self-supervised manner.
Similarly, we approximate the degradation to generate LF counterparts as:
\begin{equation}
    \mathbf{y} = \mathbf{A}\mathbf{x} + \mathbf{n}, \quad \text{where} \quad \mathbf{A} = Blur\big(DS_k(\Gamma_{\gamma}(\mathbf{x}))\big),
\end{equation}
where \(\mathbf{A} \in \mathbb{R}^{M \times N}\) represents the degradation operator, which is generally non-invertible, \(Blur, DS_{k}, \Gamma_{\gamma}\) denote Gaussian blur kernel, downsampling with a factor of $k$ and gamma transform with the coefficient $\gamma$, respectively, and \(\mathbf{n}\) denotes measurement noise. To approximate the inverse of \(\mathbf{A}\), a deep neural network \(f_\theta\) is trained to reconstruct the HF image as \(\hat{\mathbf{x}} = f_\theta(\mathbf{y})\).

However, the ill-posed nature of this inverse problem can induce \emph{hallucinations}, spurious structures not present in the ground truth \(\mathbf{x}_{\text{true}}\). 
To assist with formulation of our hypothesis, we categorize hallucinations into either \emph{Intrinsic} or \emph{Extrinsic} \cite{hallucination_tomo,videohallucer}, characterized by: 
\begin{equation}
    \mathbf{A} \hat{\bm{x}} \;\neq\; \mathbf{A}\mathbf{x}_{\text{true}},
    \hspace{1cm} (\mathbf{I} - \mathbf{A}^{+}\mathbf{A}) \hat{\mathbf{x}}
    \;\neq\;
    (\mathbf{I} - \mathbf{A}^{+}\mathbf{A}) \mathbf{x}_{\text{true}},
\end{equation}
The first term represents intrinsic hallucinations, where the reconstructed image $\hat{\mathbf{x}}$ violates data consistency, meaning its projection onto the measurement space differs from $\mathbf{A} \mathbf{x}_{\text{true}}$.  
The second term captures extrinsic hallucinations, where errors in the null space of $\mathbf{A}$ introduce structures that do not exist in $\mathbf{x}_{\text{true}}$ but appear in the reconstruction.
These hallucinations may obscure or mimic clinically significant details, potentially leading to misdiagnosis.

\subsection{Hypothesis and Justification}
\label{sec:hypothesis}
We hypothesize that combining a diffusion model, trained on a diverse dataset of HF MR images, with a conditional model can mitigate both intrinsic and extrinsic hallucinations. While conditional models excel at generating HF-like predictions from LF inputs, they are prone to hallucinations, particularly in out-of-distribution scenarios. In contrast, diffusion models provide a robust prior by learning the full distribution of plausible HF MR images rather than explicitly mapping LF to HF. This broader representation, combined with data consistency enforcement, helps to suppress errors introduced by conditional models. Below, we provide a theoretical justification of how this approach reduces hallucinations.

During the reverse diffusion process, DynamicDPS reduces hallucinations from the conditional model by leveraging: 
\begin{itemize}
    \item \textbf{Data Prior:} The diffusion model learns from a broad HF MRI distribution, mitigating \emph{extrinsic} hallucinations from incomplete measurements.
    \item \textbf{Data Consistency:} A correction term enforces alignment with $y$, mitigating \emph{intrinsic} hallucinations.
\end{itemize}

As in \cite{chung2023diffusion}, using the trained score function, the gradient of the posterior log-density at each time step is formulated as:
\begin{equation}
\nabla_{x_t} \log p_t(x_t | y) \simeq s_{\theta}(x_t, t) - \rho_{t} \nabla_{x_t} \| y - A(\hat{x}_0) \|^2_2,
\label{equ:dps}
\end{equation}
where $s_{\theta}(x_t, t)$ is the learned score function and $\rho_{t} > 0$ is the step size. 

This process can be interpreted through the lens of Maximum a Posteriori (MAP) estimation, where the goal is to find the most probable high-quality image given the measurement, yielding a MAP-like refinement trajectory grounded in Bayesian inference. 
Consequently, extrinsic hallucinations are reduced by the diffusion model's prior, $s_{\theta}(x_t, t)$, while intrinsic ones are addressed through the data consistency term, $\nabla_{x_t} \| y - A(\hat{x}_0) \|^2_2$, systematically reducing both hallucinations in the conditional model's prediction.



\subsection{Phase I: Conditional Model Prediction}  
In the first phase, any conditional IQT model can be used to predict a HF-like image \(\hat{x}_{cond}\) from the LF input \(\mathbf{y}\). While this output may still contain hallucinations, it serves as a strong initialization for the next step. Our framework is model-agnostic and can integrate with any conditional model.  

DynamicDPS reduces sampling steps by starting the reverse process from an intermediate time point. To balance efficiency and fidelity, stronger hallucinations require to initiate at earlier (noisier) stages~\cite{localdiff}. Incorrect time selection may push the conditional prediction outside the model's distribution, degrading performance. To address this, we introduce Data-Consistency-Aware Time Selection (DCATS), which optimizes the starting time point \(t_{\text{optimal}}\) per sample.  

DCATS begins by constructing a memory bank that stores, for each time step $t$, the average data likelihood computed over a reference dataset separate from the test set. During testing, the reference data likelihood is compared to the scaled likelihood of the conditional model’s prediction, and $t_{optimal}$ is selected as the time step $t$ that minimizes the discrepancy between the two:  
\begin{equation}
\label{eq:forward}
t_{\text{optimal}} 
\,=\, \underset{t}{\arg\min} 
\;\Bigl|\, 
p\bigl(\mathbf{y} \mid \hat{\mathbf{x}}_{\text{cond}}\bigr)^{\tau}
\;-\; 
\mathbb{E}_{\text{ref,t}}\bigl[p\bigl(\mathbf{y}_{\text{ref}} \mid \hat{\mathbf{x}}_{\text{ref}, 0|t}\bigr)\bigr]
\Bigr|.
\end{equation}

Here, \(p(\mathbf{y} \,|\, \hat{\mathbf{x}}_{\text{cond}})^{\tau}\) represents the likelihood of the test measurement given the predicted image from conditional models, scaled with the temperature hyper-parameter, $\tau = 0.4$, and \(p(\mathbf{y_{\text{ref}}} \,|\, \hat{\mathbf{x}}_{\text{ref}, 0|t})\) denotes the likelihood of a set of reference data at time step \(t\). By selecting \(t_{\text{optimal}}\) to minimize this likelihood mismatch, the method accelerates sampling while maintaining image quality, achieving a significant improvement over standard DPS.

\subsection{Phase II: Diffusion Model}  
In the second phase, pre-trained diffusion models on HF MR scans handle hallucinations by enforcing data consistency at each time step to reduce intrinsic hallucinations. Meanwhile, the diffusion prior corrects extrinsic hallucinations by guiding the reconstruction in the null space of the measurement operator, ensuring plausibility where measurements provide no constraint.

After computing \( t_{\text{optimal}} \), we use \( \hat{x}_{cond} \) from Phase I as a warm start for the diffusion process, significantly reducing sampling steps. 
Since most time steps are skipped, computational resources are focused on optimizing the data consistency term to suppress intrinsic hallucinations effectively.  

\subsubsection{Data Consistency Loss Function}
Vanilla DPS uses only an $\ell_2$ penalty, $\|\mathbf{y} - A(\hat{\mathbf{x}}_{\theta})\|_{2}^{2}$, which can be suboptimal for noisy or heavily undersampled inputs. To address this, we add two auxiliary terms, resulting in the total data consistency loss, $L_{\mathrm{DC}}$ as:

\begin{equation}
\label{eq:dataconsistency}
L_{\mathrm{DC}} 
= \|\mathbf{y} - A(\hat{\mathbf{x}}_{\theta})\|_{2}^{2}
+ \lambda_{1}\,\mathrm{Edge}\!\bigl(\mathbf{y},\, A(\hat{\mathbf{x}}_{\theta})\bigr)
+ \lambda_{2}\,\mathrm{SSIM}\!\bigl(\mathbf{y},\, A(\hat{\mathbf{x}}_{\theta})\bigr),
\end{equation}

where $\mathrm{Edge}$ and $\mathrm{SSIM}$ denote $\ell_2$-Sobel Edge and $\ell_2$-SSIM losses, respectively. The loss weights $\lambda_1$ and $\lambda_2$ are set to 0.5 and 0.1. These terms preserve anatomical boundaries and enhance local structural similarity.

\subsubsection{Dynamic Step-Size Optimization via Wolfe's Line Search}
\label{sec:wolfe_line_search}
A key limitation of data-consistency-based diffusion models is sensitivity to fixed step sizes, which can be sub-optimal for each test input to reduce intrinsic hallucination effectively under a limited number of iterations. In our reverse diffusion process, we adopt \textbf{Wolfe's line search}~\cite{wolf} to dynamically select the step size \(\alpha_t\) for the update of $x_{t}$ and define $\phi(\alpha)$ as:
\begin{equation}
    \mathbf{x}_{t+1} \leftarrow \mathbf{x}_t + \alpha_t \,\mathbf{p}_t,
    \hspace{1cm}
    \phi(\alpha) = f(\mathbf{x}_t + \alpha\,\mathbf{p}_t),
\end{equation}
where $\mathbf{p}_t$ is a direction (e.g., negative gradient of an objective) and $f(\cdot)$ includes both the data fidelity term and diffusion prior. Wolfe's line search optimizes $\alpha_t$ satisfying 1. \textbf{Armijo rule}, $\phi(\alpha) \le \phi(0) + c_1 \,\alpha \,\phi'(0)$, ensuring a non-trivial decrease in $f(\cdot)$, and 2. \textbf{Curvature condition}, $|\phi'(\alpha)| \le c_2 \,|\phi'(0)|$, preventing overshooting.
Here, $0 < c_1 < c_2 < 1$, and $\phi'(0) = \nabla f(\mathbf{x}_t)^\mathsf{T} \mathbf{p}_t$. These conditions can be interpreted as finding an upper and lower bounds of step size, which allows faster reduction of intrinsic hallucination without overshooting.

\section{Experiments}
\subsection{Experimental Setup}
\subsubsection{Datasets}
We evaluate our method on the \textit{Human Connectome Project (HCP)} dataset \cite{hcp}. The simulated low-field (LF) test set (600 images) is split into: (1) \textbf{In-distribution}: LF images at $2.8 mm$ resolution with $\gamma=$ 0.7, matching training conditions; (2) \textbf{Out-of-distribution}: lower contrast ($\gamma=$ 0.4) or lower spatial resolution ($4.2 mm$). For real MRI, LF T1-weighted scans were acquired on a 0.36T MagSense 360 scanner with non-isotropic voxels ($1.0 \times 1.0 \times 7.2mm^3$, slice thickness: $6.0 mm$, gap: $1.2 mm$), and the corresponding HF image ($1.0 \times 1.0 \times 1.0 mm^3$ isotropic) was registered with the LF for visual comparison.

\subsubsection{Baselines \& Metrics}  
We evaluate our approach against four baselines: U-Net~\cite{unet}, ESRGAN~\cite{ESRGAN}, and DPS~\cite{chung2023diffusion}.  
DPS and DynamicDPS use the same score-matching model pre-trained on the HCP dataset for a fair comparison.  
Methods requiring test-time parameter optimization are excluded.  
For quantitative evaluation, we use PSNR (Peak Signal-to-Noise Ratio), SSIM (Structural Similarity Index), and LPIPS~\cite{lpips} (Learned Perceptual Image Patch Similarity).  

\subsection{Main Results}

Table~\ref{table:results1} shows a quantitative comparison against baselines. While conditional models perform similarly to diffusion-based methods on IND data, their performance degrades on OOD, underscoring their sensitivity to data shifts. 
In contrast, all of our approaches perform robustly on OOD and outperform baselines. 
Notably, \textit{Ours with ESRGAN} on OOD data boosts PSNR and LPIPS by over 50\%, demonstrating the ability to refine the output under a few time steps. 
We also include \textit{Ours} as an ablation, using only DynamicDPS without any conditional backbone, which still outperforms DPS on both IND and OOD, validating the benefit of our dynamic data consistency formulation.

Efficiency is assessed by comparing inference speed on IND data. Although DynamicDPS is computationally heavy due to the data consistency step, leveraging the time-skipping strategy via DCATS speeds up by over 80\% compared to DPS~\cite{chung2023diffusion}, using only on average 50 time steps with U-Net vs. 1000 for DPS.
\begin{table}[!tb]
  \label{results}
  \centering
  \footnotesize
  \begin{tabular}{cccccccc}
    \toprule
      & \multicolumn{3}{c}{In-distribution} &\multicolumn{3}{c}{Out-of-distribution} & {Inf. time}\\
       & \textit{PSNR} ($\uparrow$) & \textit{SSIM} ($\uparrow$) & \textit{LPIPS} ($\downarrow$) & \textit{PSNR} ($\uparrow$) & \textit{SSIM} ($\uparrow$) & \textit{LPIPS} ($\downarrow$) & \textit{(s)}\\
    \midrule
    U\cite{unet} & 28$\pm$.97 & .87$\pm$.03 & \textbf.10$\pm$.02 & 24$\pm$1.9 & .80$\pm$.06 & .18$\pm$.05 & -\\
    G\cite{ESRGAN} & 27$\pm$.02 & .85$\pm$.02 & \textbf{.09}$\pm$.01 & 18$\pm$5.6 & .76$\pm$.04 & .24$\pm$.03 & -\\
    DPS\cite{chung2023diffusion} & 27$\pm$.92 & .84$\pm$.03 & .10$\pm$.02 & 26$\pm$1.9 & .81$\pm$.06 & .14$\pm$.04 & 196$\pm$0.6\\
    \myrowcolor%
    Ours& \textbf{29}$\pm$1.0 & .88$\pm$.03 & .10$\pm$.02 & \textbf{27}$\pm$1.9 & \textbf{.86}$\pm$.06 & \textbf{.13}$\pm$.03 & -\\
    \myrowcolor%
    Ours$+$U & \textbf{29}$\pm$1.0 & \textbf{.89}$\pm$.03 & \textbf{.09}$\pm$.02 & 26$\pm$1.9 & \textbf{.86}$\pm$.06 & \textbf{.13}$\pm$.04 & \textbf{37}$\pm$0.8\\
    \myrowcolor%
    Ours$+$G & \textbf{29}$\pm$.90 & \textbf{.89}$\pm$.03 & \textbf{.09}$\pm$.02 & 26$\pm$.06 & \textbf{.86}$\pm$.06 & \textbf{.13}$\pm$.03 & 38$\pm$0.8\\
    \bottomrule
  \end{tabular}
  \caption{Quantitative comparison of image quality across in/out-of-distribution datasets, where an upward arrow indicates that a higher value is better. "Ours" refers to DynamicDPS with diffusion models only (T=1000). "U" and "G" denote U-Net~\cite{unet} and ESRGAN~\cite{ESRGAN}, respectively.}
  \label{table:results1}
\end{table}

\begin{figure}[!tbp]
\centering
\includegraphics[width=1.0\textwidth]{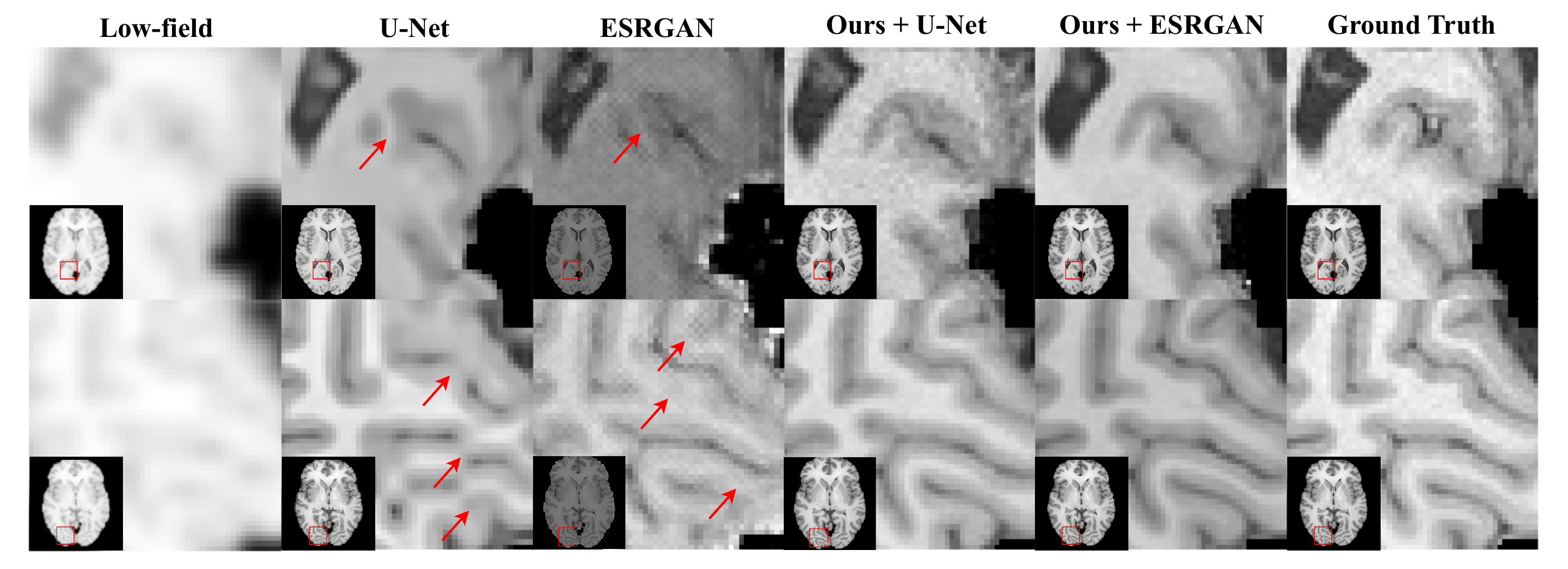}
\caption{Visual comparisons on OOD data. Zoomed-in regions are marked with red boxes, and red arrows indicate hallucinated areas.}
\label{fig:results_main}
\end{figure}

Figure~\ref{fig:results_main} presents a qualitative comparisons against baseline conditional models. As indicated by the red arrows, conditional models generate both intrinsic and extrinsic hallucinations such as false sulci and contrast artifacts. In contrast, DynamicDPS consistently reduces hallucinations while preserving anatomical structures. These results highlight the effectiveness of our approach in visually also mitigating hallucinations across various conditional models.

\subsection{Further Analysis}
\begin{figure}[!tbp]
     \centering
     \begin{subfigure}[b]{0.47\textwidth}
         \centering
         \includegraphics[width=\textwidth]{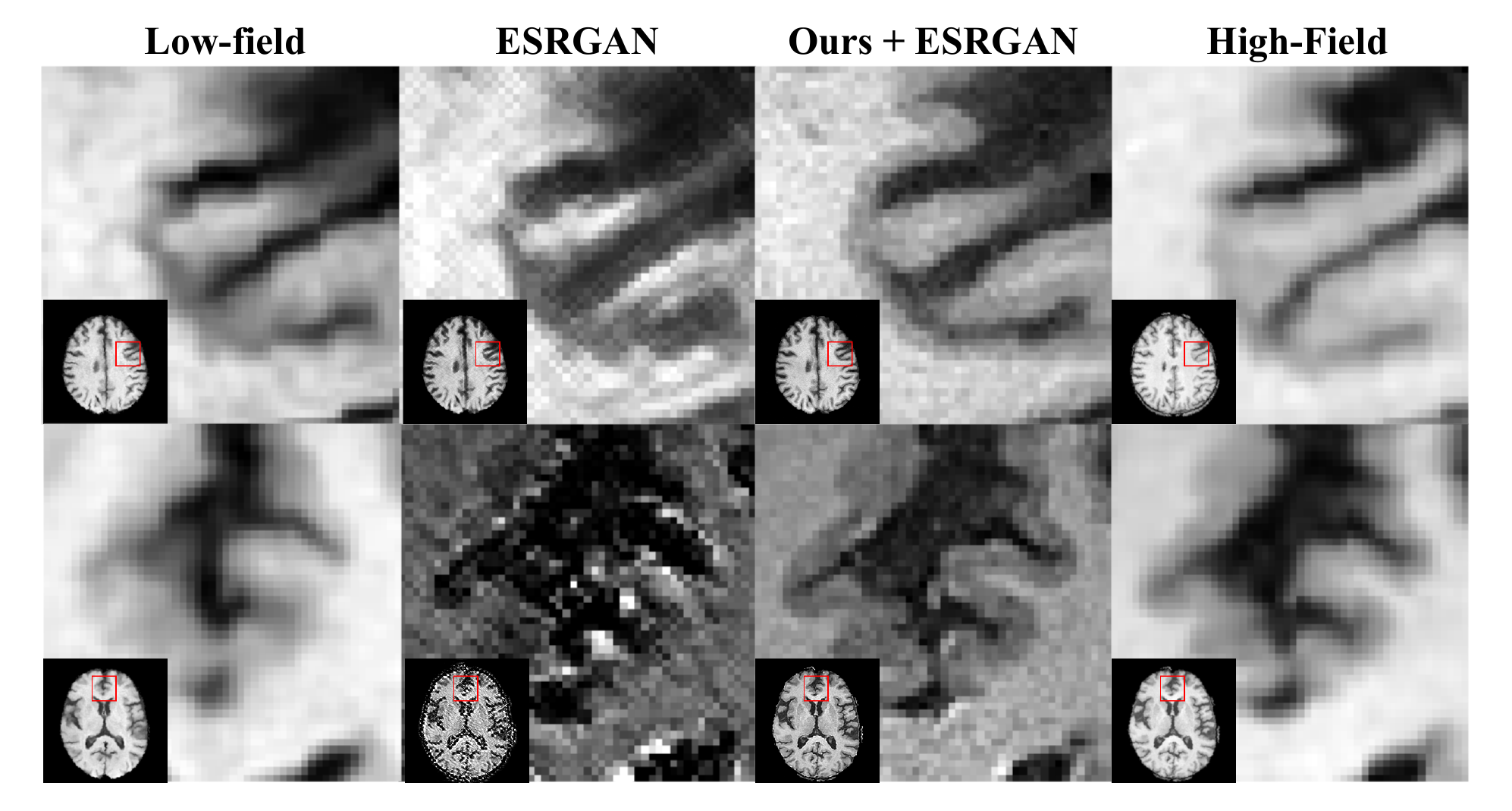}
         \caption{Visual results each conditioned on real low-field (top) and high-field (bottom) MR scans.}
         \label{fig:results_real}
     \end{subfigure}
     \hfill
     \begin{subfigure}[b]{0.49\textwidth}
         \centering
         \includegraphics[width=\textwidth]{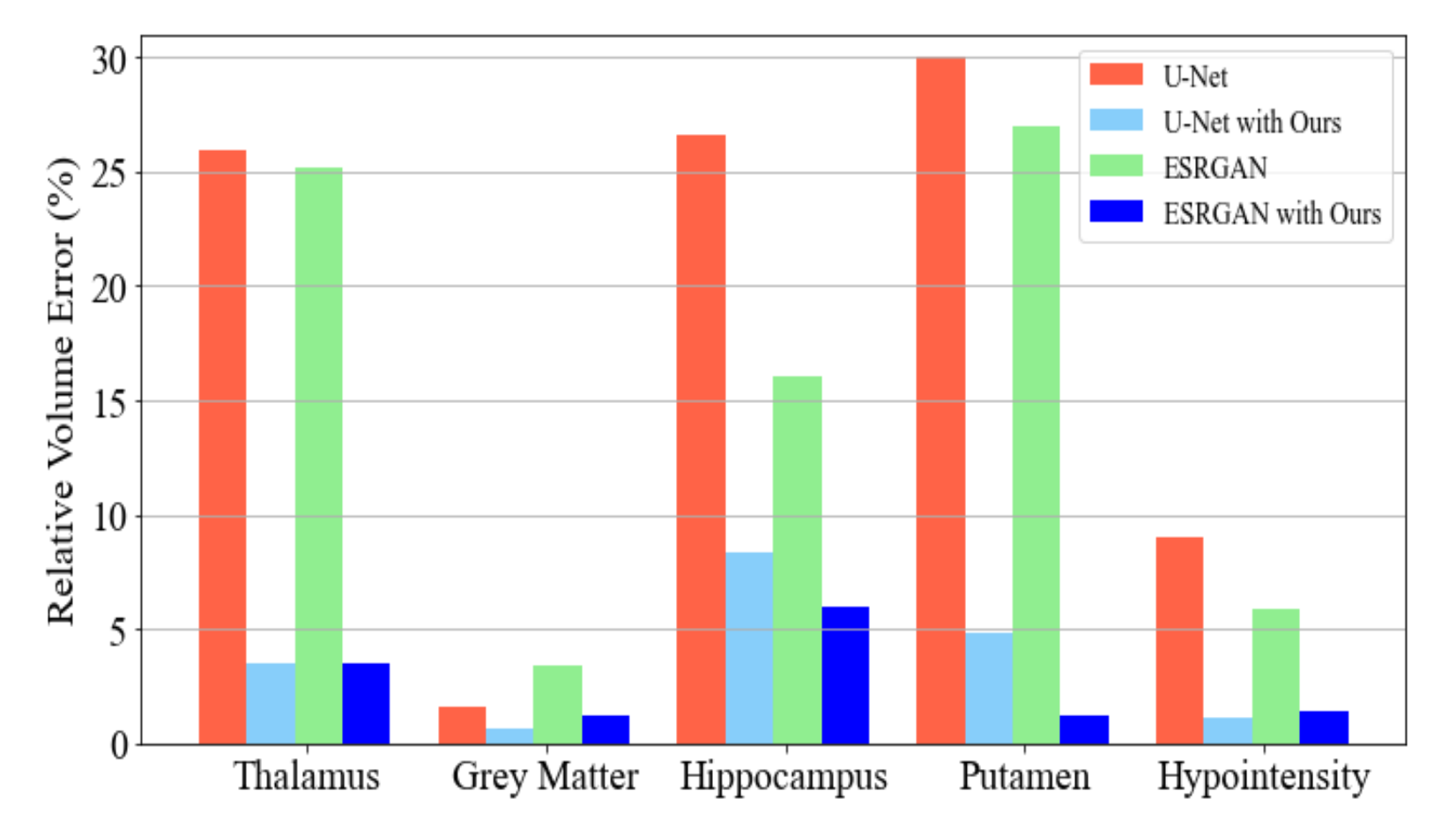}
         \caption{Relative volume error for brain tissues using FastSurfer~\cite{fastsurfer}}
         \label{fig:furthera}
     \end{subfigure}
    \caption{Comparisons against baselines on (a) real low-field and high-field MR scans with zoomed-in regions are marked with red boxes, (b) volume estimation for hallucination evaluation.}
    \label{fig:further}
\end{figure}

Figure~\ref{fig:results_real} shows visual results tested on real LF and HF scans. While ESRGAN struggles to enhance contrast and fails on the HF scan, introducing extrinsic hallucinations, ours improves tissue contrast without visible hallucinations in both scans. Notably, even with \emph{minimal tuning of the data consistency parameter}, our approach achieves superior visual contrast and image quality than the HF scan (second row), underscoring its robustness across varying degradation levels.

Global image quality metrics (e.g. PSNR, SSIM) often fail to capture hallucinations, as they prioritize global image fidelity and thus under-penalize such localized features. Instead, we estimated brain tissue volumes using FastSurfer~\cite{fastsurfer} and computed the relative volumetric error (RVE) following~\cite{iqt_stochastic}, defined as $RVE = \frac{2|V_{pred} - V_{gt}|}{V_{pred} + V_{gt}}$, where $V$ denotes the volume of a tissue. This metric suits hallucination evaluation, as hallucinations can distort tissue boundaries and volume estimates. Figure~\ref{fig:furthera} shows that our method significantly reduces volume error by more than 15\% in critical tissues such as the thalamus, hippocampus, putamen, and hypointensity across all conditional models. 
These structures are prone to hallucination due to their small, variable size and lower SNR compared to the cortex in some sequences, yet crucial for applications such as Alzheimer's diagnosis. 
While hallucination metrics~\cite{hallucination_tomo,hallucination_metric} were discussed earlier, we opted volume estimation to provide a clinically relevant assessment due to the unavailability of implementations. In future work, we plan to incorporate these established metrics into our evaluation framework.

\section{Conclusion}
We propose DynamicDPS, a diffusion-based framework that reduces hallucinations by integrating conditional predictions with measurement-driven refinement. By selecting an optimal starting point via DCATS and applying Wolfe's line search-based data consistency, it dynamically enforces data consistency during sampling. Experiments on synthetic and real MR scans demonstrate its effectiveness.
Future work will explore automatic parameter estimation and spatially adaptive step-size weighting for improved hallucination reduction. Leveraging a pre-trained diffusion model on high-quality medical data, DynamicDPS offers a fine-tuning-free, model-agnostic solution broadly applicable to low-field MRI, advancing the democratization of reliable medical imaging.

\begin{credits}
\subsubsection{\ackname} This work is supported by the EPSRC-funded UCL Centre for Doctoral Training in Intelligent, Integrated Imaging in Healthcare (i4health) under grant number EP/S021930/1. 
The work of DCA, MF, and HG is supported by the Wellcome Trust (award 221915/Z/20/Z), the MRC (award MR/W031566/1), and the NIHR UCLH Biomedical Research Centre.
\subsubsection{\discintname} The authors have no competing interests to declare that are relevant to the content of this article.
\end{credits}
%
%

\bibliographystyle{splncs04}
\bibliography{Paper-4526}





\end{document}